\documentclass[twocolumn,showpacs,preprintnumbers,aps,nofootinbib]{revtex4-1}

\usepackage{graphicx}  
\usepackage{dcolumn}
\usepackage{bm}
\usepackage{amsmath,amssymb,amsfonts}
\usepackage{latexsym}
\usepackage{color}

\def\nn    {\nonumber}

\begin{document}

\title{\boldmath
Constraining $t \to u$ flavor changing neutral Higgs coupling at the LHC}

\author{Wei-Shu Hou, Ting-Hsiang Hsu and Tanmoy Modak}
\affiliation{Department of Physics, National Taiwan University, Taipei 10617, Taiwan}
\bigskip

\begin{abstract}
We study the constraints on $t\to u$ flavor changing neutral Higgs (FCNH) coupling,
and how it may be explored further at the Large Hadron Collider (LHC). 
In the general two Higgs doublet model, such transitions 
can be induced by a nonzero $\rho_{tu}$ Yukawa coupling. 
We show that such couplings can be constrained by 
existing searches at the LHC for
$m_H$, $m_A$ and, $m_{H^+}$ in the sub-TeV range, 
where $H$, $A$ and $H^+$ are the exotic $CP$-even, $CP$-odd and charged scalars. 
We find that a dedicated $ug\to t H/tA \to t t \bar u$ search can probe 
the available parameter space of $\rho_{tu}$ down to
a few percent level for $200\,\mbox{GeV} \lesssim m_H,\,m_A \lesssim 600$\;GeV, 
with discovery possible at high luminosity.
Effects of how other extra top Yukawa couplings, such as $\rho_{tc}$ and $\rho_{tt}$,
dilute the sensitivity of the $\rho_{tu}$ probe are discussed.
\end{abstract}

\maketitle

\section{Introduction}\label{intro}

The 125 GeV scalar boson $h$, only discovered~\cite{h125_discovery} in 2012, 
combines with the longitudinal components of the massive vector bosons 
to form the weak scalar doublet of the Standard Model (SM).
But with one scalar doublet established naturally brings in 
the question of a second doublet, i.e. the so-called~\cite{Branco:2011iw} 
two Higgs doublet model (2HDM).
Although it is popular~\cite{Branco:2011iw} to use a discrete symmetry
to impose ``Natural Flavor Conservation''~\cite{Glashow:1976nt}
so all ``dangerous'' flavor changing neutral Higgs (FCNH) couplings 
are removed, it is also well known that this may not be necessary~\cite{Branco:2011iw}.
Indeed, upon the discovery of $h$, the $t \to ch$ decay~\cite{Hou:1991un} 
search was advocated~\cite{Chen:2013qta} and quickly
pursued by ATLAS~\cite{Aad:2014dya} at the LHC, 
and further efforts are recorded~\cite{PDG} by the Particle Data Group ({PDG}).
As another example, CMS saw early on with 8 TeV data 
some hint~\cite{Khachatryan:2015kon} for $h \to \tau\mu$ decay.
Though it subsequently disappeared~\cite{PDG},
it did bring about considerable interest in FCNH couplings.

As elucidated in Ref.~\cite{Chen:2013qta}, the $t \to ch$ decay 
occurs via the $c_\gamma \rho_{tc}$ coupling,
where $c_\gamma \equiv \cos\gamma$ is the mixing angle of
$h$ with the $CP$-even scalar boson $H$ of the exotic doublet,
which is the one that carries the FCNH $\rho_{tc}$ coupling.
With subsequent Higgs property 
studies~\cite{Khachatryan:2016vau,Sirunyan:2018koj,Aad:2019mbh}, 
it became clear that $h$ resembles very closely the Higgs boson of SM,
and the $h$--$H$ mixing angle $c_\gamma$ seems rather small.
This may be the reason behind the non-observation~\cite{PDG} of $t \to ch$ 
so far, without implying $\rho_{tc}$ to be small.
Demonstrating~\cite{Hou:2017hiw} that there is 
quite some parameter space for $c_\gamma$ to be small in the 2HDM context, 
it was advocated that mass-mixing hierarchy suppression~\cite{Hou:1991un} 
of FCNH couplings involving lighter generation fermions, 
augmented by the smallness of $c_\gamma$ (``alignment''), 
can explain the absence of low energy FCNH effects 
without the need to invoke NFC.  
Thus, extra Yukawa couplings are rather general in the 2HDM setting 
and should be pursued experimentally, and not just at the LHC.
The ``Model III'' of Ref.~\cite{Hou:1991un} was therefore elevated
to the general 2HDM (g2HDM), even promoted~\cite{Chang:2017wpl} 
as a possible future ``SM2'', the SM with two Higgs doublets.

\begin{figure}[b]
\center
\includegraphics[width=0.35\textwidth]{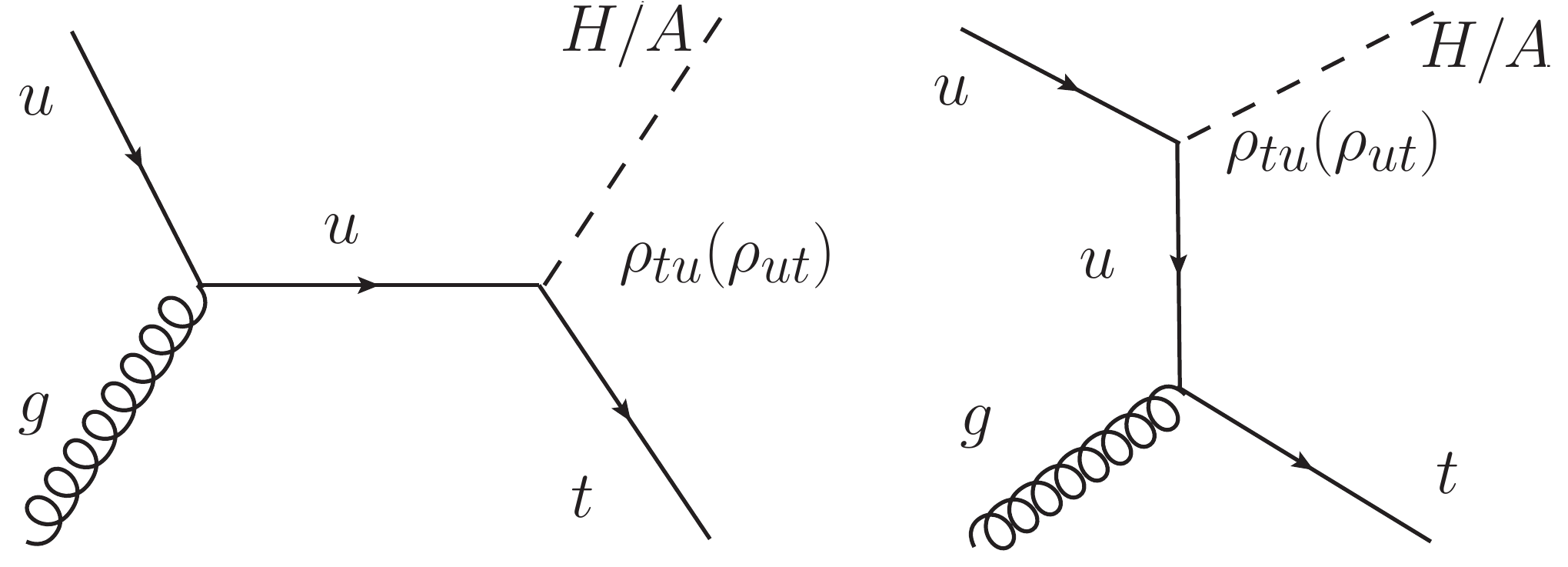}
\caption{Feynman diagrams for $ug\to tH/tA$.}
 \label{feyndiag}
\end{figure}

{Having introduced the g2HDM, we write down the couplings of 
the $CP$-even scalars $h$, $H$ and $CP$-odd scalar $A$ to fermions  
as~\cite{Chen:2013qta,Hou:2017hiw,Davidson:2005cw}
\begin{align}
\mathcal{L} = 
& -\frac{1}{\sqrt{2}} \sum_{f = u, d, \ell}
     \bar f_{i} \Big[(-\lambda^f_{ij} s_\gamma + \rho^f_{ij} c_\gamma) h  \nn\\
+ & (\lambda^f_{ij} c_\gamma + \rho^f_{ij} s_\gamma ) H
    - i\,{\rm sgn}(Q_f) \rho^f_{ij} A\Big]  R\, f_{j}
 +{\rm H.c.},
 \label{eff}
\end{align}
where $L,R\equiv (1\mp\gamma_5)/2$,
 $i,j = 1, 2, 3$ are generation indices and summed over,
$c_\gamma = \cos\gamma$ and $s_\gamma = \sin\gamma$, 
and $u$, $d$, $\ell$ are up and down type quarks and charged leptons, respectively. 
The matrices $\lambda^f_{ij}\; \equiv \sqrt{2}m_i^f \delta_{ij}/v$ are 
the usual Yukawa couplings related to mass in SM,
whereas $\rho^f_{ij}$ are in general non-diagonal and complex.
We do not consider $H^+$ effects in this work,
but consider $\rho_{tu}$-induced processes at the LHC,
including $ug \to tH/tA$ production (see Fig.~\ref{feyndiag}).
We refrain from quoting the Higgs potential for g2HDM here.
Instead, we treat the scalar boson masses as parameters,
but state that we have checked that they satisfy
the usual requirements of perturbativity, positivity and unitarity,
as well as other constraints such as electroweak oblique parameters 
(see e.g. Refs.~\cite{Hou:2019qqi,Hou:2019mve,Ghosh:2019exx}).}

In the experimental pursuit of $t \to ch$,
one actually searches for $t \to ch$, $uh$ simultaneously.
It turns out that the bound on $t\to uh$ is not better than $t \to ch$, 
i.e. the current 95\% C.L. bound from ATLAS~\cite{Aaboud:2018oqm} gives
\begin{align}
{\cal B}(t \to uh)  < 1.2 \times 10^{-3}, \ \, 
{\cal B}(t \to ch)  < 1.1 \times 10^{-3}, 
 \label{tuh-ATL}
\end{align}
based on 36.1\,fb$^{-1}$ data at 13~TeV, which is better than 
the CMS result~\cite{Sirunyan:2017uae} based on similar amount of data. 
This may seem surprising since single top production via $\rho_{tu}$ is taken into account.
One may think that $\rho_{tu}$ should naturally be much smaller than $\rho_{tc}$, 
{but this is not based on our current experimental knowledge.}
It was pointed out~\cite{Hou:2019uxa} that
$B \to \mu\bar\nu$ decay probes the $\rho_{tu}\rho_{\tau\mu}$ product.
The process will be pursued by Belle~II~\cite{Kou:2018nap},
where a deviation of the ratio
 ${\cal R}_B^{\mu/\tau} = {\cal B}(B \to \mu\bar\nu)/{\cal B}(B \to \tau\bar\nu)$
from the SM expectation of 0.0045 would indicate~\cite{Hou:2019uxa} 
nonzero $\rho_{tu}$ in g2HDM.
What can LHC do to check $\rho_{tu} \neq 0$?
{In this paper we focus on $ug \to tH/tA \to tt\bar u$ production,
i.e. $ug \to tH/tA$ (see Fig.~\ref{feyndiag}) followed by $H/A \to t\bar u$, 
leading to same-sign top signature.}

In  the next section we first summarize the constraint on 
$\rho_{tu}$ from searches at the LHC, including $t\bar tt\bar t$ search. 
We turn to $ug\to tH/tA\to t t \bar u$ (conjugate process always implied
{unless specified}) in Sec.~III 
and use it to constrain or discover the $\rho_{tu}$ coupling~{\cite{rho_ut}. 
We focus on $m_A$,\,$m_H \in (200,\,600)$\;GeV, which is 
 allowed in g2HDM~\cite{Hou:2019qqi,Hou:2019mve,Ghosh:2019exx}. 
Heavier $m_A,\,m_H$ are possible, but discovery prospect 
is reduced due to rapid fall off in parton luminosities.
As the $\rho_{tc}$-induced $cg\to tH/tA\to t t \bar c$ process~\cite{Hou:1997pm,Kohda:2017fkn,Hou:2018zmg,Hou:2019gpn} 
 (see also
 Refs.~\cite{Altmannshofer:2016zrn,Iguro:2017ysu,Iguro:2018qzf,Cao:2019qrb}) 
can be  misidentified as $ug\to tH/tA\to tt\bar u$ 
due to inefficient $c$-jet tagging, we outline 
a procedure to distinguish between the two processes. 
We comment briefly on the effect of the diagonal $\rho_{tt}$ coupling
in Sec.~IV, before offering our conclusion.

\section{\boldmath Current Constraints on $\rho_{tu}$}
 \label{constr}

As stated, our actual knowledge of the strength of $\rho_{tu}$ 
is actually quite poor.

The $h$ boson couples to $tu$ as $c_\gamma \rho_{tu}$, 
hence $\mathcal{B}(t\to u h)$ search constrains 
$\rho_{tu}$ coupling for finite $c_\gamma$.
The latest ATLAS result based on 36.1 fb$^{-1}$ data at 13 TeV sets 
the 95\% C.L. limit $\mathcal{B}(t\to u h) < 1.1\times 10^{-3}$~\cite{Aaboud:2018oqm}, 
as given in Eq.~(\ref{tuh-ATL}), which is better than 
the CMS limit~\cite{Sirunyan:2017uae} of 
$\mathcal{B}(t\to u h) < 4.7 \times 10^{-3}$ based on 35.9 fb$^{-1}$. 
We illustrate the ATLAS limit~\cite{Aaboud:2018oqm} in Fig.~\ref{ttouh} 
as the blue shaded region in the $c_\gamma$--$\rho_{tu}$ plane, 
while the weaker CMS limit is not displayed.
Taking $c_\gamma = 0.2$ as example, one gets
$|\rho_{tu}| \lesssim 0.5$ at $95\%$ C.L., which is 
rather weak, and weakens further for smaller $c_\gamma$.

Stronger constraints on $\rho_{tu}$ arise from the $t\bar tt\bar t$, 
or $4t$ search, which does not depend on $c_\gamma$.
Let us first focus on the CMS $4t$ search, which is based on 137~fb$^{-1}$ at 13 TeV, 
i.e. with full Run~2 data~\cite{Sirunyan:2019wxt}, more than three times 
the data size of the preceding study~\cite{Sirunyan:2017roi}.
Depending on the number of charged leptons ($e$, $\mu$) and 
$b$-tagged jets, the search in Ref.~\cite{Sirunyan:2019wxt} is 
divided into several signal regions (SRs) and two control regions (CRs), 
with the baseline selection criterion of at least two same-sign leptons.
We find that the most stringent constraint on $\rho_{tu}$ 
arises from the control region of $t\bar t W$, 
which is denoted as CRW~\cite{Sirunyan:2019wxt}.
Induced by the $\rho_{tu}$ coupling, the $ug\to tH/tA \to t t \bar u$ process 
would contribute to this CRW.

\begin{figure}[t]
\center
\includegraphics[width=.345 \textwidth]{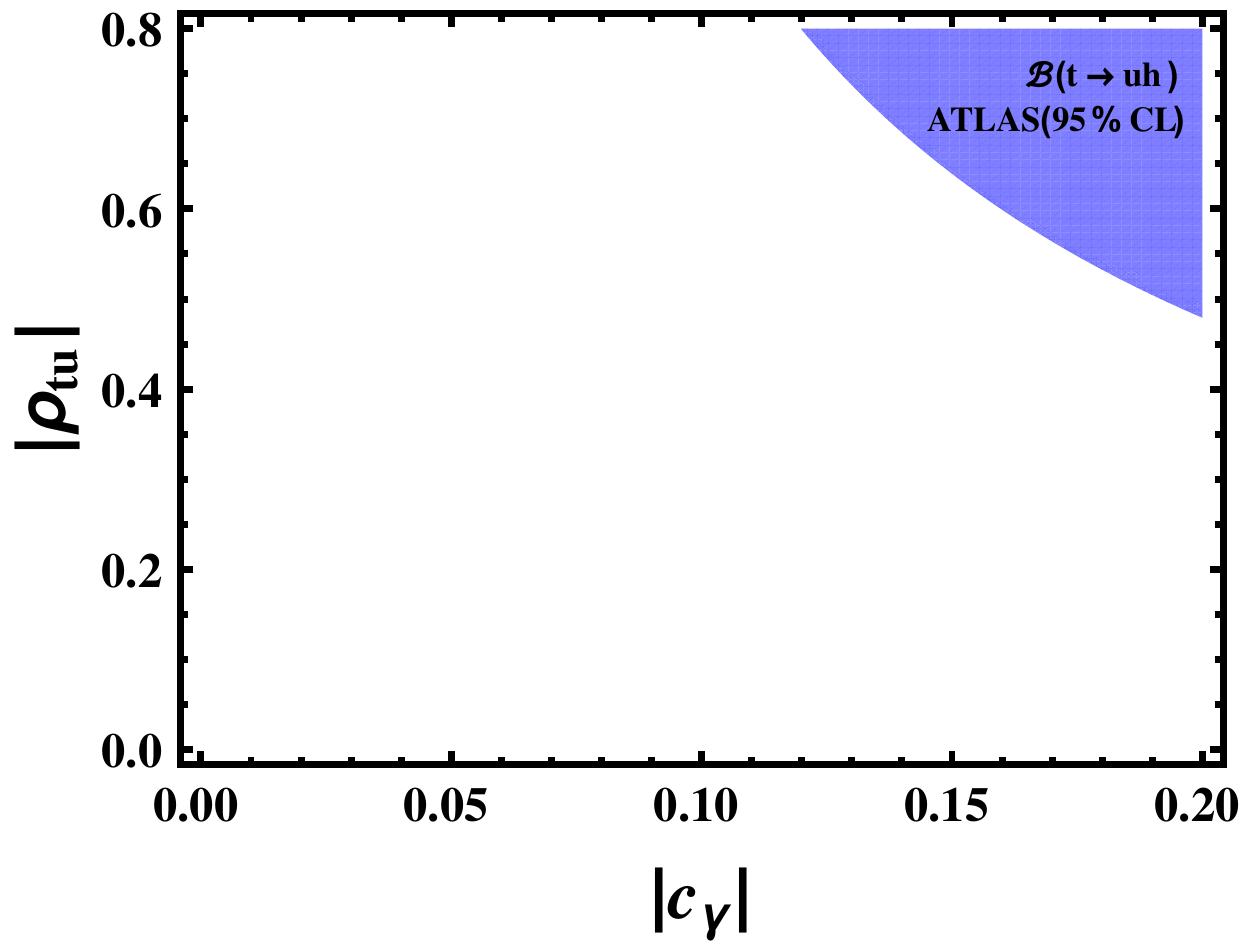}
\caption{
$\mathcal{B}(t\to uh)$ constraint in $|c_\gamma|$--$|\rho_{tu}|$ plane. } 
\label{ttouh}
\end{figure}

CRW of the CMS $4t$ search~\cite{Sirunyan:2019wxt} is defined as 
containing two same-sign leptons plus two to five jets with two $b$-tagged.
The selection cuts are as follows.
Leading\,(subleading) lepton transverse momentum 
should satisfy $p_T$\,$>$\,$25$\,(20)\;GeV. 
The pseudorapidity of electrons (muons) should satisfy
 $|\eta| < 2.5$\;($2.4$), while all jets satisfy $|\eta| < 2.4$.
The events are selected if $p_T$ of ($b$-)jets satisfy 
any of the following three conditions~\cite{info-Jack}:
(i) both $b$-jets satisfy $p_T$\,$>$\,$40$\;GeV;
(ii) one $b$-jet with $p_T$\,$>$\,$20$\;GeV and $20$\,$<$\,$p_T$\,$<$\,$40$\;GeV
 for the second $b$-jet,
 with $p_T$\,$>$\,$40$\;GeV for the third jet;
(iii) both $b$-jets satisfy $20$\,$<$\,$p_T$\,$<$\,$40$\;GeV, with two extra jets 
 each satisfying $p_T$\,$>$\,$40$\;GeV.
$H_T$, defined as the scalar sum of $p_T$ of all jets, 
should satisfy $H_T$\,$>$\,$300$\;GeV, while $p_T^{\rm miss}$\,$>$\,$ 50$\;GeV.
To reduce the Drell-Yan background with a charge-misidentified electron,
events with same-sign electron pairs with $m_{ee}$\,$<$\,$12$\;GeV are rejected.
With these selection cuts, CMS reports 338 observed events in CRW,
while the expected total number of events 
(SM backgrounds plus $4t$) is at $335 \pm 18$~\cite{Sirunyan:2019wxt}.

\begin{figure*}[htbp]
\center
\includegraphics[width=.39 \textwidth]{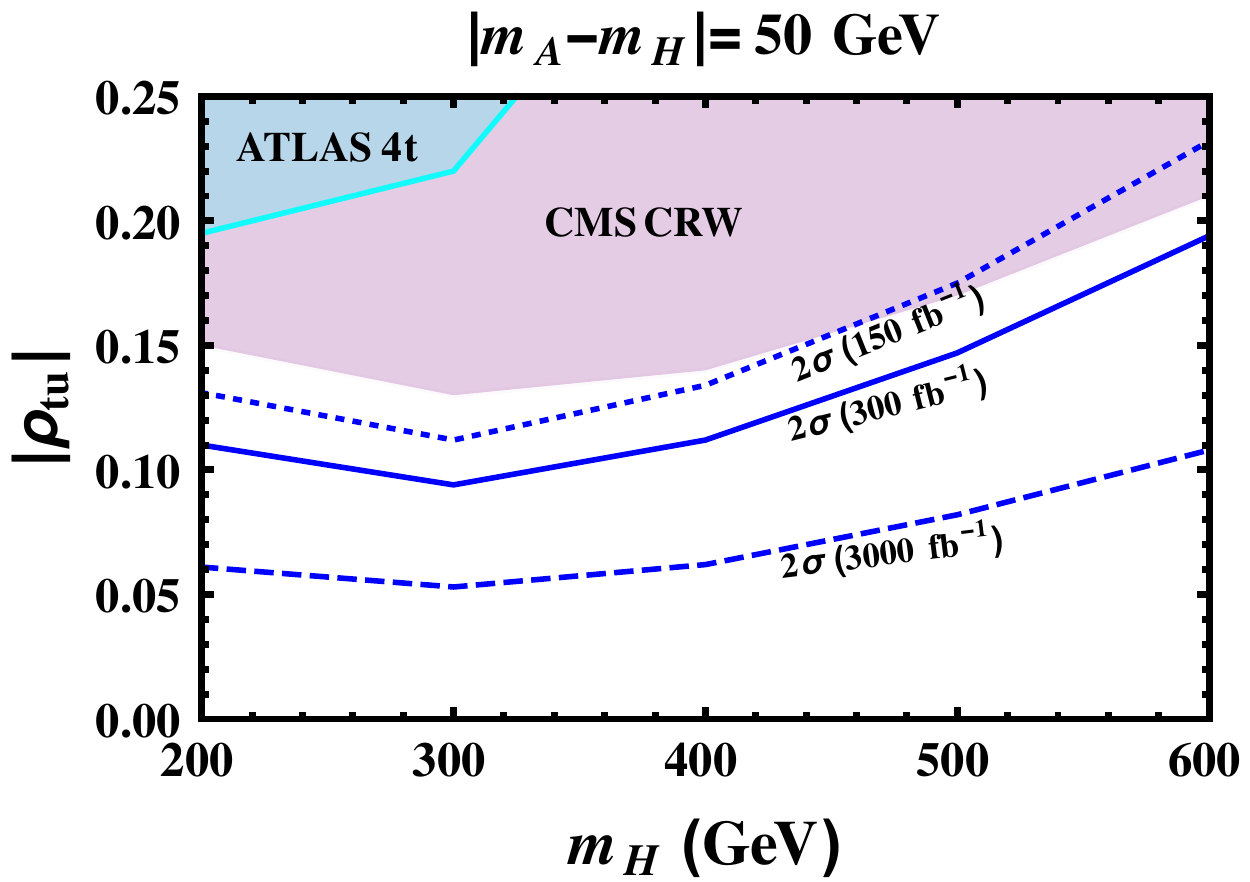}
\includegraphics[width=.39 \textwidth]{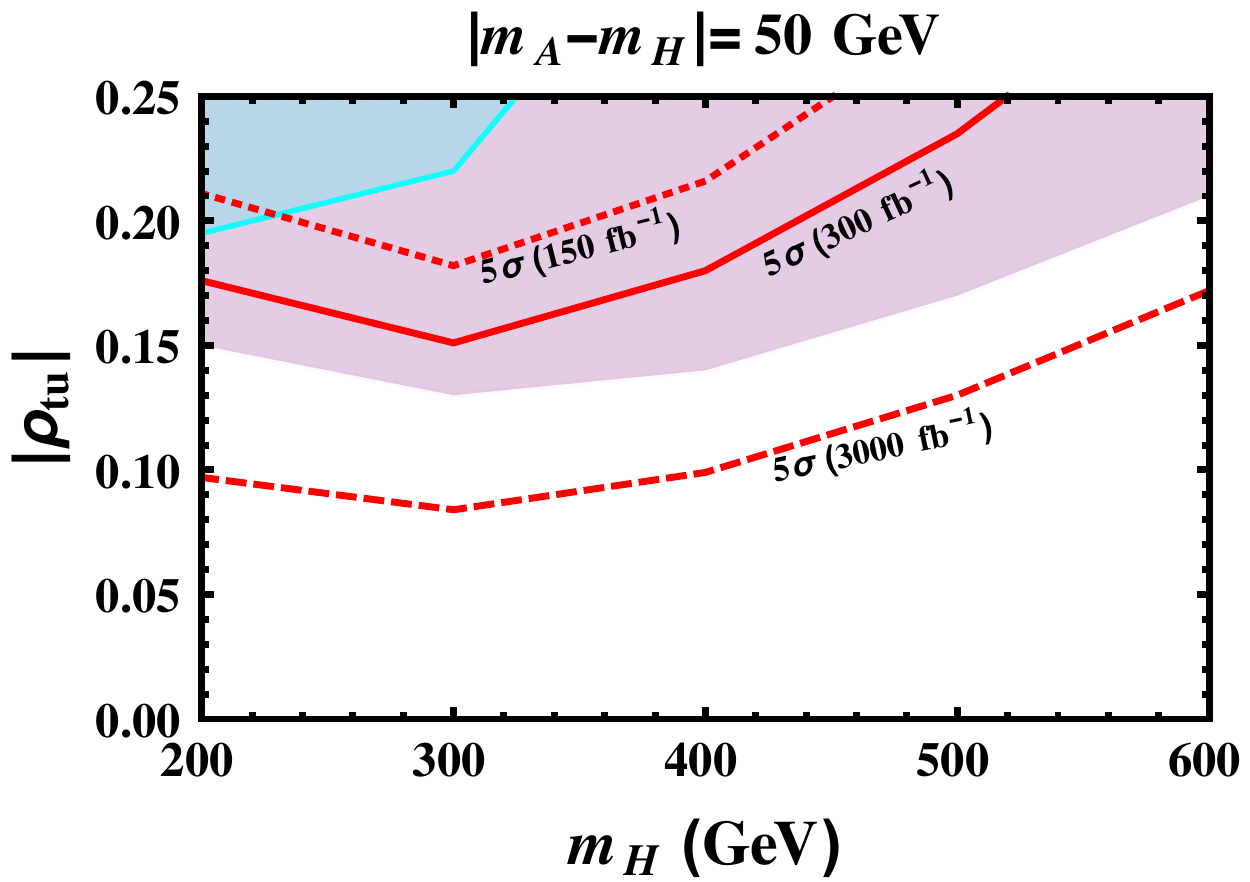}
\caption{
Exclusion limits  [left] and discovery reaches [right] 
for $|\rho_{tu}|$ by the same-sign top signature with various integrated luminosities at the 14 TeV LHC, where the purple and cyan regions are excluded respectively 
by CMS CRW~\cite{Sirunyan:2019wxt} 
and ATLAS CRttW2$\ell$~\cite{ATLAS:2020hrf} control regions.
See text for details.
}
 \label{exclu}
\end{figure*}

To calculate our limits, we generate signal events using 
MadGraph5\_aMC@NLO~\cite{Alwall:2014hca}
 (denoted as MadGraph5\_aMC) at leading order (LO) 
with default parton distribution function (PDF) set NN23LO1~\cite{Ball:2013hta}, 
{interface} with PYTHIA~6.4~\cite{Sjostrand:2006za}
 for showering and hadronization, 
and MLM matching~\cite{Alwall:2007fs} prescription
 for matrix element and parton shower merging.
The event samples are then fed into Delphes~3.4.2~\cite{deFavereau:2013fsa}
 for fast detector simulation, where we follow
 the CMS-based detector analysis for CRW, 
 utilize the default $b$-tagging efficiency and light-jet rejection,
 with jets reconstructed via anti-$k_T$ algorithm. 
The effective model is implemented in FeynRules~\cite{Alloul:2013bka}.

The $\rho_{tu}$-induced process  $pp\to tH/tA \to t t \bar u$ 
(non-resonant  $ug \to t t \bar u$ and $t$-channel $H/A$ 
exchange $uu\to tt$ processes are included) 
with both top quarks decaying semileptonically 
contributes to CRW of CMS $4t$ search. 
Setting all other $\rho_{ij} = 0$, we estimate 
the contribution 
for $\rho_{tu}=1$ and then scale the cross section by $|\rho_{tu}|^2$, 
assuming narrow $H/A$ widths with $\mathcal{B}(H/A\to t \bar u) = 50\%$. 
We then demand that the sum of the number of events expected from SM 
and those from $\rho_{tu}$-induced processes agree with the observed 
number of events within $2\sigma$ uncertainty of expectations. 
We display the $2\sigma$ exclusion limits obtained via CRW 
in Fig.~\ref{exclu} as the purple shaded regions, 
where we assume Gaussian behavior for simplicity.
That is, we simplify and do not follow the more 
precise estimation~\cite{Cowan:2010js} of exclusion limits 
using likelihood function with Poisson counting.

ATLAS has also searched for $4t$ production~\cite{ATLAS:2020hrf}
with 139~fb$^{-1}$, but categorizing into different SRs and CRs. 
Again, the CR for $t\bar t W$, called $\rm{CRttW2\ell}$, is the most relevant. 
It is defined as
 at least two same-sign leptons ($e^\pm \mu^\pm$ or $\mu^\pm \mu^\pm$),
 plus at least four jets with at least two $b$-tagged. 
The same-sign leptons are required to have $p_T > 28$\;GeV
 with $|\eta^\mu| < 2.5$ and $|\eta^e| < 1.5$. 
All jets should satisfy $p_T > 25$\;GeV and $|\eta| < 2.5$.
If the number of $b$-jets is equal to two, 
{or the number of $b$-jets is $\geq 3$ but with no more than 5 jets,}
 the scalar $p_T$ sum over all jets and same-sign leptons
 (note the difference 
  in definition from CMS), $H_T$, should satisfy {$H_T < 500$\;GeV}. 
Unlike CRW for CMS, ATLAS 
does not give the observed number of events in $\rm{CRttW2\ell}$,
but provides a figure of comparison between data and prediction 
in the variable $\sum p_T^\ell$ (see Ref.~\cite{ATLAS:2020hrf} for definition).
We extract~\cite{extrac} from this figure the number of expected and observed events 
for $\rm{CRttW2\ell}$, finding $378\pm 10$ and $380$, respectively,
where we have simply added the errors in quadrature for the expected events 
from each $\sum p_T^\ell$ bin.

To extract the constraint, we follow the same event selection procedure
as before, but use the ATLAS-based detector card of Delphes.
Assuming that the number of events for $pp\to tH/tA \to t t \bar u$ 
plus SM stay within $2\sigma$ of the expected number of events, 
we illustrate the exclusion limits from ATLAS $\rm{CRttW2\ell}$ 
by the cyan shaded regions in Fig.~\ref{exclu}. 
Mainly due to differences in selection cuts, the ATLAS constraint on $\rho_{tu}$ is weaker.
From CMS $4t$ search we find 
{$\rho_{tu} \lesssim 0.13$--$0.15$}
is still allowed for 
{$200\,\mbox{GeV}\lesssim m_H \lesssim 400$\;GeV}, 
while slightly larger values open up for $m_H > 400$\;GeV.
In this vein, we stress that we have illustrated for 
$|m_H - m_A| = 50$\;GeV, as there is strong cancellation between
$ug \to tH \to tt\bar c$ and $ug \to tA \to tt\bar c$ amplitudes
for $H,\,A$ {that are nearly degenerate in mass and width}.

We remark that supersymmetry search in similar event topologies 
can in principle constrain $\rho_{tu}$.
However, such analyses now typically require $H_T$ and/or missing energy 
that are too large for our purpose.
The selection criteria could be relaxed with $R$-parity violation, 
e.g. the ATLAS search~\cite{Aad:2019ftg} for squark pair production,
but the selection cuts are still too strong to give meaningful constraint. 
We note further that the ATLAS search for new phenomena 
in events~\cite{Aaboud:2018xpj} with same-sign dileptons and $b$-jets
(36.1\;fb$^{-1}$ at 13\;TeV) has similar SRs, 
but the cuts are again strong and the selection criteria different, 
such that it does not give relevant constraint for our study.

\section{\boldmath Same-sign top signature from $\rho_{tu}$}
 \label{coll}

Even though the existing CMS $4t$ search with full LHC Run~2 data
can set meaningful constraints on $\rho_{tu}$,
it is not optimized for $ug \to tH/tA \to tt\bar u$ search.
In this section, we perform a dedicated study of 
the $ug \to tH/tA \to tt\bar u$ process at the LHC, 
targeting exclusion or discovery of a stand-alone $\rho_{tu}$ coupling.

\subsection{Discovery and Exclusion Limits}

The $pp\to tH/ tA + X \to t t\bar u + X$ process can be 
searched for in events containing
 same-sign dilepton ($ee$, $\mu\mu$, $e\mu$),
 at least three jets with at least two $b$-tagged and one non-$b$-tagged jet,
 plus $E_T^{\rm{miss}}$, which we denote as same-sign top.
The final state topology will also receive contribution from $uu \to tt$ 
via $t$-channel {$A/H$} exchange which we include as signal. 
The dominant backgrounds are $t\bar t Z$, $t\bar t W$, $4t$ and $t\bar t h$,
 while $3t + W$, $3t + j$ and $tZ +$\,jets are subdominant. 
In addition,  if the lepton charge gets misidentified (charge- or $Q$-flip), 
with the misidentification efficiency at
$2.2\times 10^{-5}$~\cite{ATLAS:2016kjm,Aaboud:2018xpj,Alvarez:2016nrz},
the $t\bar t+$\,jets and $Z/\gamma^*+$\,jets processes would also contribute.
We remark that the CMS study~\cite{Sirunyan:2017uyt} 
with similar final state topology but with slightly different cuts 
finds the ``nonprompt'' backgrounds at $\sim 1.5$ times 
the $t\bar{t}W$ background, which is significant.
As the nonprompt backgrounds are not properly modeled in Monte Carlo simulations, 
we simply add this component to the overall background 
at 1.5 times the $t\bar t W$ background after selection cuts.

\begin{figure*}[htbp!]
\center
\includegraphics[width=.37 \textwidth]{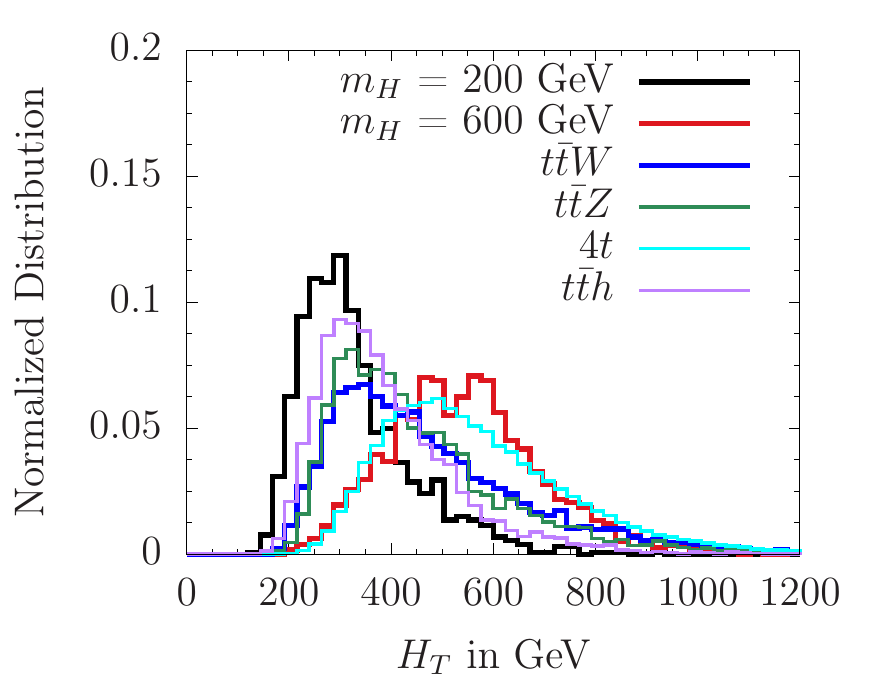}
\includegraphics[width=.37 \textwidth]{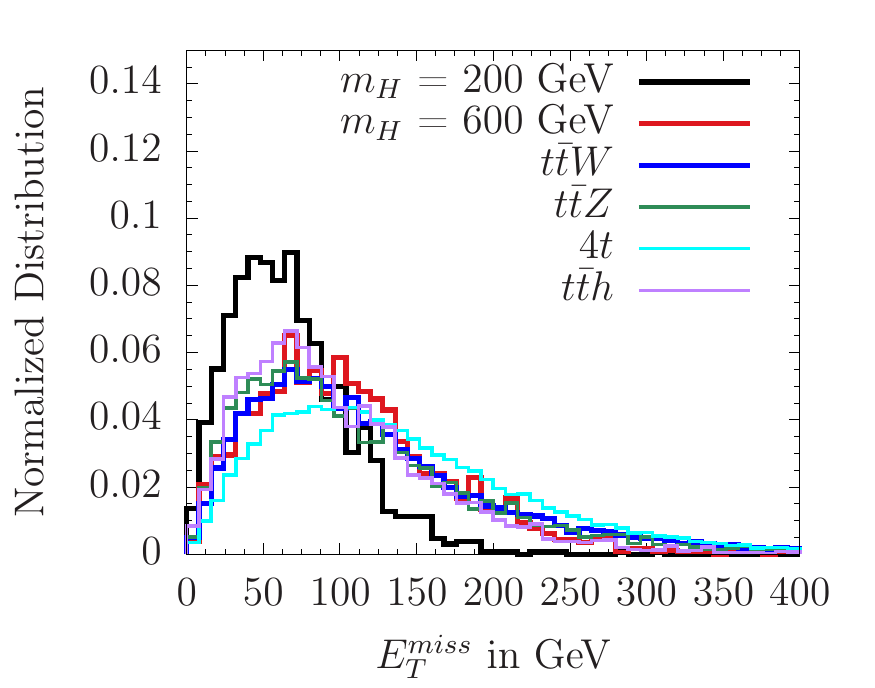}
\caption{
The normalized $H_T$ (left) and $E^{\rm miss}_{T}$ (right) distributions for the signal and leading backgrounds.
See text for details.
}\label{kindist}
\end{figure*}

We generate signal and background events as in the previous section at LO 
via MadGraph5\_aMC for $\sqrt{s}=14$\;TeV, follow the same 
showering, hadronization and ME, and parton shower merging and matching.
We adopt here the default ATLAS-based detector card of Delphes.
The LO  $t\bar{t} W^-$\;($t\bar{t} W^+$), $t\bar t Z$, 
$4t$, $t\bar t h$ and, $tZ+$\,jets  cross sections  are normalized to 
next-to-leading order $K$ factors 
1.35 (1.27)~\cite{Campbell:2012dh}, 
2.04~\cite{Alwall:2014hca}, 
1.44~\cite{Alwall:2014hca}, 
1.27~\cite{twikittbarh}, and 
1.56~\cite{Campbell:2013yla}, respectively.
We assume the same $K$ factor for $\bar{t}Z+$\,jets background for simplicity.
The $Q$-flip $Z/\gamma^*+$\,jets and $t\bar t +$\,jets backgrounds are 
corrected to next-to-next-to-leading (NNLO) order cross sections by 
$1.27$~\cite{Hou:2017ozb} and 
$1.84$~\cite{twiki}, respectively. 
We utilize FEWZ 3.1~\cite{Li:2012wna} to obtain 
the NNLO factor for $Z/\gamma^*+$\,jets background. 
The signal cross sections and $3t + W$, $3t + j$\;backgrounds are kept at LO.

\begin{table}[b]
\centering
\begin{tabular}{c |c | c c c }
\hline\hline
\,$m_H$ [$\Gamma_H$] (GeV) & \,$m_A$ [$\Gamma_H$] (GeV)  &\,cross section (fb)  \\
\hline 
                       200 [0.81]                 &   250 [4.14]                                       & 18.9                \\
                       300 [8.07]                 &   350 [12.0]                                      & 25.6               \\
                       400 [15.7]                 &   450 [19.6]                                       & 18.1                    \\
                       500 [23.2]                 &   500 [26.7]                                       & 10.6                \\
                       600 [30.2]                 &   650 [33.6]                                       & \ 6.0               \\
\hline\hline
\end{tabular}
\caption{
Mass and width of $H$ and $A$ for $\rho_{tu}=1$, 
and same-sign top signal cross section at 14 TeV after selection cuts.}
 \label{sig}
\end{table}

\begin{table}[b]
\centering
\begin{tabular}{c |c c c c }
\hline
\hline
                     \,backgrounds                 & \,cross section (fb)      \\
\hline 
                       $t\bar{t}W$                & 1.31               \\
                       $t\bar{t}Z$                 & \,\ 0.264               \\
                       $4t$                            & \,\ 0.092               \\
                       $t\bar t h$                  & \,\ 0.058                \\                       
                       $Q$-flip                    & \,\ 0.024                 \\
                       $tZ+$\,jets                 & \,\ 0.007                \\
                       $3t+W$                     & \,\ 0.001                \\
                       $3t+j$                       & \,\ \ \,0.0004           \\
\hline
\hline
\end{tabular}
\caption{
Background cross sections after selection cuts.}
 \label{backg}
\end{table}

To reduce backgrounds, we follow a cut based analysis that is 
different from CRW of CMS $4t$ search, 
and optimize for $pp\to t A/tH + X \to t t\bar u + X$ as follows.
The leading (subleading) lepton should have $p_T > 25$ (20)\;GeV,
while $|\eta| < 2.5$ for both leptons. 
All three jets should satisfy $p_T> 20$\;GeV and $|\eta| < 2.5$. 
The missing energy in each event should satisfy $E^{\rm miss}_{T} > 30$\;GeV.
The separation $\Delta R$ between a lepton and any jets ($\Delta R_{\ell j}$), 
between the two $b$-jets ($\Delta R_{bb}$), 
and between any two leptons ($\Delta R_{\ell\ell}$)
should all satisfy $\Delta R > 0.4$. 
We finally demand that selected events should satisfy $H_T > 300$\;GeV,
where $H_T$ is defined according to ATLAS,i.e. 
including the $p_T$ of the two leading leptons. 

We plot in Fig.~\ref{kindist} the normalized $H_T$ and $E^{\rm miss}_{T}$ 
distributions before selection cuts for signal and dominant backgrounds. 
For signal we choose the two representative $m_H = 200$ and 600\;GeV values 
(with $m_A = m_H+50$\;GeV) for illustration.
The signal cross section for different $m_H$ with $|m_A-m_H| = 50$\;GeV 
and background cross sections after the selection cuts 
are summarized in Tables~\ref{sig} and~\ref{backg}, respectively. 
We have assumed $m_H$ to be lighter than $m_A$.

To estimate the exclusion limit ($2\sigma$) and discovery potential ($5\sigma$), 
we utilize the test statistics~\cite{Cowan:2010js}
\begin{align}
Z(x|n)=\sqrt{-2\ln\frac{L(x|n)}{L(n|n)}},\label{poissn}
\end{align}
where $L(x|n) =  e^{-x}x^n/n!$ is the likelihood function of Poisson probabilities 
with $n$ the observed number of events, and $x$ is either the number of events predicted 
by the background-only hypothesis $b$, or signal plus background hypothesis $s+b$.
For exclusion ($s+b$ hypothesis) 
we demand $Z(s+b|b) \geq 2$ for $2\sigma$,
while for discovery ($b$ hypothesis) $Z(b|s+b) \geq 5$ for $5\sigma$.
Utilizing the signal cross sections for the 
reference $\rho_{tu}=1$ value in Table~\ref{sig} and
the background cross sections in Table~\ref{backg}, 
we find the exclusion and discovery contours 
in $m_H$--$\rho_{tu}$ plane (with $m_A = m_H+50$ GeV) 
for different integrated luminosities in the left and right 
panels of Fig.~\ref{exclu}, respectively,
where we have interpolated the contours for $m_H$ values
other than the ones given in Table~\ref{sig} for simplicity.

The exclusion and discovery contours are plotted in Fig.~\ref{exclu} 
as blue and red lines, respectively, for the three different integrated 
luminosities of 150 (dotted), 300 (solid) and 3000\;fb$^{-1}$ (dashed).
The 150\;fb$^{-1}$ data size reflects the target luminosity for Run~2,
but the contours are generated with $\sqrt{s}=14$\;TeV rather than 13\;TeV. 
We find that, with 150 (300)\;fb$^{-1}$ one could exclude
$|\rho_{tu}|\gtrsim 0.13~(0.11)$ if 
$200\,\mbox{GeV} \lesssim m_H \lesssim\,400$\;GeV, 
whereas $|\rho_{tu}|\gtrsim 0.18$--0.25~(0.15--0.19) 
for $400\,\mbox{GeV}\lesssim m_{H}\lesssim 600$\;GeV.
With full High Luminosity LHC (HL-LHC) data, 
i.e. with 3000\;fb$^{-1}$, 
the exclusion limit can reach down to $|\rho_{tu}| \gtrsim 0.06$
 for $m_H \lesssim 400$\;GeV, and $|\rho_{tu}|\,{\gtrsim}\,0.1$
 for 400\,{GeV}\,$\lesssim m_H \lesssim 600$\;GeV.
One would need larger $|\rho_{tu}|$ for discovery. 
For example, {the discovery contours for 150 and 300\;fb$^{-1}$ 
lie in the regions excluded by CMS CRW.} 
For the HL-LHC dataset, $|\rho_{tu}|\sim 0.1~(0.17)$ would be 
sufficient for discovery for $200\,\mbox{GeV}\lesssim m_H \lesssim 400$\;GeV 
($400\,\mbox{GeV}\lesssim m_H \lesssim 600$\;GeV).

\subsection{Distinguishing $\rho_{tu}$ and $\rho_{tc}$ Effects}

Unless the final state charm can be efficiently tagged (which is not the case), 
the $cg \to tH/tA \to tt\bar c$ processes also give rise to the same-sign top 
signature for nonzero $\rho_{tc}$. In this subsection, we outline a procedure
to distinguish same-sign top signatures induced by $\rho_{tu}$ vs $\rho_{tc}$.

The valence $u$-quark induced $ug \to tH/tA \to tt\bar u$ process has much larger
cross section compared to $\bar u g \to \bar tH/\bar t A \to \bar t \bar t u$. 
So one expects the former to be considerably larger than the latter.  
To understand the relative significance of $ug \to tH/tA \to tt\bar u$, 
we take a benchmark point with {$\rho_{tu}=0.13$}, 
$m_H,\,m_A = 300,\,350$\;GeV that is still allowed by Fig.~\ref{exclu}. 
To distinguish between the signature induced by $\rho_{tu}$ vs $\rho_{tc}$,
we separate positively charged vs negatively charged same-sign dileptons. 
Following the same analysis as in the previous subsection, 
we find the signal (background) cross sections at $\sqrt{s}=14$\;TeV 
for the ++ and $--$ charged dileptons to be {0.5}\;fb and {0.06}\;fb
($\sim2.35$\;fb and $\sim 1.38$\;fb), respectively.
We then find the significance 
for dileptons with ++ charge to be {$\sim 4.1\sigma$} ({$\sim 13 \sigma$}) 
with 300 (3000)\;fb$^{-1}$, while the corresponding significance for 
$--$ charged dileptons is at {$\sim 1\sigma$} ({$\sim\,2.7\sigma$}). 
Note that the former (latter) arises from the
 $ug \to tH/tA \to tt\bar u$ ($\bar ug \to tH/tA \to tt\bar u$) process. 
Thus, separating the ++ from $--$ same-sign dilepton events, 
one expects the ++ dileptons to emerge earlier than the $--$.
We have again assumed the non-prompt background to be $\sim 1.5$ 
times the $t\bar t W$ background, while $Q$-flip background is assumed
at half the value given in Table~\ref{backg} 
for the respective signatures.

In comparison, the $c$-quark induced $cg \to tH/tA \to tt\bar c$ and 
$\bar c$-quark induced $\bar c g \to \bar tH/\bar t A \to \bar t \bar t c$ 
processes should have similar cross sections. 
Assuming all $\rho_{ij}=0$ except $\rho_{tc}$ we find, for example, that
$\rho_{tc}=0.36$ is allowed at $2\sigma$ by CRW of CMS $4t$ search 
for $m_H,\,m_A = 300$, $350$\;GeV.
Following 
the same cut based analysis for these parameter values, 
we find the cross sections at $\sqrt{s}=14$~TeV for ++ and $--$ charged dilepton 
processes at 0.074 and 0.081\;fb, respectively,
which translates to $\sim$\;2.7$\sigma$ ($\sim 8.4\sigma$)
 and $\sim3.8\sigma$ ($\sim 11.9\sigma$)
 with 300 (3000)\;fb$^{-1}$ integrated luminosity. 
That is, both ++ and $--$ same-sign dilepton events are at similar level, 
which contrasts with the $\rho_{tu}$-induced same-sign dilepton events. 

So far we have discussed scenarios when 
either $\rho_{tu}$ or $\rho_{tc}$ is nonzero.
Recasting the results from Ref.~\cite{Crivellin:2013wna},  it was found~\cite{Altunkaynak:2015twa} that $|\rho_{tu}^*\rho_{tc}| \gtrsim 0.02$ 
is excluded by $D$--$\overline{D}$ mixing for 
$m_H \approx m_A \approx m_{H^\pm} \simeq 500$\;GeV,
 which would be even more stringent for lighter exotic scalars. 
This gives the ballpark of the constraint when 
both $\rho_{tu}$ and $\rho_{tc}$ are nonzero.
%
A detailed analysis treating both $\rho_{tu}$ and $\rho_{tc}$ nonzero 
would be studied elsewhere.

\section{Discussion and Outlook}

Let us comment on the impact of turning on $\rho_{tt}$.
As $\rho_{tt} \neq 0$ would induce $H/A\to t \bar t$ decays, 
the $4t$ search constraints from CRW of CMS and 
$\rm{CRttW2\ell}$ of ATLAS would weaken for $m_H\,(m_A) > 2 m_t$ 
due to $\mathcal{B}(H/A\to t \bar t) \neq 0$. 
In particular, $\rho_{tt}=0.5$ is still allowed for 
$m_H,\,m_A,\,m_{H^\pm} \sim 200$--600\;GeV~\cite{Ghosh:2019exx}. 
For $\rho_{tu} = 0.15$ and $\rho_{tt} = 0.5$, 
{$\mathcal{B}(H/A\to t \bar u + \bar t u)$ would be suppressed 
by $\sim 70\%$--90\%} for $400\,\mbox{GeV} \lesssim m_H \lesssim 600$\,GeV, 
weakening the limits from CRW of CMS $4t$ search. 
Nonzero $\rho_{tu}$ and $\rho_{tt}$ may 
also induce $ug \to tH/tA  \to t t \bar t $ (triple-top) 
and $ug\to b H^+\to b t \bar b$ signatures, where the latter process 
may even emerge from Run~2 data~\cite{Ghosh:2019exx}. 
Such final states can also arise from $\rho_{tc}$ coupling. 
However, separating ++ and $--$ same-sign dileptons 
can in principle differentiate between $\rho_{tu}$ and $\rho_{tc}$ couplings. 
Scenarios when $\rho_{tu}$, $\rho_{tc}$ and $\rho_{tt}$ are all nonzero 
would receive multiple constraints, in particular from flavor physics. 
A study involving all three couplings is beyond the scope of this work. 
{However, based on the extensive work on
$ug,\,cg \to tH/tA  \to tt\bar u,\,tt\bar c,\,t t \bar t$ processes reported
or cited here, we advertise a public twiki page~\cite{twiki_FCNH} 
where interested LHC workers could use to join the quest.}

At this point, it is useful to recall that $\rho_{tt}$ provides a robust 
driver~\cite{Fuyuto:2017ewj} for electroweak baryogenesis (EWBG) in g2HDM, 
even for $|\rho_{tt}|$ values at the percent level, which provides strong motivation.
If $\rho_{tt}$ is sizable, it would make probing nonzero $\rho_{tu}$ 
more challenging at the LHC.
However, we have emphasized our current experimental knowledge,
and such knowledge on $\rho_{tu}$ comes primarily from the LHC at present.
Even if one takes EWBG into consideration, we note 
a second, backup mechanism~\cite{Fuyuto:2017ewj}:
$\rho_{tc}$ at ${\cal O}(1)$ with near maximal phase can also drive EWBG
{if $\rho_{tt}$ accidentally vanishes in g2HDM.
%
However, it would still make probing $\rho_{tu}\neq 0$
rather challenging, and the LHC experiments would have to try their
best at the HL-LHC, as we have tried to illustrate.
This is especially so if ${\cal B}(B \to \mu\bar\nu)/{\cal B}(B \to \mu\bar\nu)$
is found by Belle~II to differ from SM expectation.
On the other hand, baryogenesis may not occur through g2HDM,
hence one should exploit the full potential of the LHC.
}

{In summary,
we pose the question:
 ``If the flavor changing neutral Higgs coupling $\rho_{tu}$ is nonzero,
 how can one check this at the LHC?'' With only $\rho_{tu} \neq 0$, 
we show that it is possible with HL-LHC, by comparing the 
significance of positively vs negatively charged same-sign dilepton events.
Nonzero $\rho_{tc}$ can mimic $\rho_{tu}$-induced events,
while $\rho_{tt} \neq 0$ would further dilute the sensitivity to finite $\rho_{tu}$.
The issue would become important if the ratio of 
$B \to \mu\bar\nu$ decay rate to $B \to \tau\bar\nu$ is found by 
Belle~II to deviate from Standard Model expectation.}

\vskip0.2cm
\noindent{\bf Acknowledgments.--} \
This work is supported by MOST\;106-2112-M-002-015-MY3,
and 108-2811-M-002-537 of Taiwan, and NTU 108L104019.


\end{document}